\documentclass{iopart}
\usepackage{hyperref}
\usepackage{amsfonts,graphicx}
\newcommand{\aob}[2]{\mathrel{\mathop{#1}\limits_{\mathop#2}}}

\begin{document}
\title[Note on transformation to general curvilinear coordinates for
Maxwell's\dots]{Note on transformation to general curvilinear
coordinates for
Maxwell's curl equations\\
(\textit{Is the magnetic field vector `axial'?})}
\author{D M Shyroki}
\address{COM-DTU, Technical University of Denmark, 2800 Kgs.~Lyngby, Denmark}
\ead{\mailto{ds@com.dtu.dk}}


\begin{abstract}
Arguments for $\mathbf H$ and $\mathbf B$ to be considered
`axial', or pseudo vectors, are revisited. As a point against, we
examine the complex-coordinate method for numerical grid
truncation and mode loss analysis proved very successful in
computational electrodynamics. This method is not compatible with
convention that $\mathbf H$ and $\mathbf B$ are axial.
\end{abstract}

\section{Introduction}

In physics, one often encounters `symmetries', i.e.\ situations
when structural fields (as the permittivity $\epsilon$ and
permeability $\mu$ in classical macroscopic electrodynamics) are
invariant under the given transformation $\mathbf x' = \mathbf
x'(\mathbf x)$. To obtain general conclusions on dynamic behavior
of such symmetric systems without solving the governing dynamic
equations, one needs to know also how the functional fields
(electric and magnetic fields in Maxwell's equations) are
transformed with $\mathbf x' = \mathbf x'(\mathbf x)$. Physical
intuition and experiment is generally what provides us with this
knowledge.

Important symmetries are with respect to coordinate reflections,
as $\mathbf x \rightarrow \mathbf x' = (-x^1,x^2,x^3)$, and
coordinate inversion, $\mathbf x \rightarrow \mathbf x' =
(-x^1,-x^2,-x^3) = -\mathbf x$. These are improper transformations
whose Jacobian, $J^{i'}_{i} = \partial x^{i'} /
\partial x^i$, has negative determinant: $\det J^{i'}_{i} = \Delta
<0$. In numerous textbooks on classical mechanics and
electrodynamics, vector-like objects are classified into polar, or
ordinary vectors, and axial, or pseudo vectors~--- depending on
their behavior under improper transformations. Polar vectors are
transformed under inversion as $\mathbf V' = -\mathbf V$, while
axial as $\mathbf W' = \mathbf W$.

The $\mathbf V' = -\mathbf V$ transformation under coordinate
inversion implies that the $\mathbf V$ vector remains unchanged in
the reference (`absolute', `physical') space. On the opposite,
$\mathbf W$ is reflected in the absolute space upon coordinate
inversion. How is it possible at all that such `unrealistic'
pseudo quantities, dependent on the way we describe them, survive
in the physical picture of reality? The answer is that they are
normally related to observable quantities through the cross
product which is defined so that it compensates for the inversion
of pseudo vectors in absolute space.

Such situation is not quite satisfactory, however. Although many
quantities in classical physics are not directly observable
(measurable) and hence their `realism' can always be questioned by
a radical empiricist, a natural trend is to put as much `realism'
into each quantity as possible. In this paper we discuss an
alternative convention for the cross product that allows to
eliminate pseudo quantities from classical physics. Interestingly,
it turns out to be more than just a matter of convention or
convenience once the complex-coordinate transformations are
concerned.

\section{Maxwell's equations}\label{sec:Maxwell}

Classical theory of electromagnetism is usually built upon the
fundament of Maxwell's equations~--- a set of dynamic field
equations which we write in standard three-vector notation as
\begin{eqnarray}
    \nabla\widetilde{^\times} \mathbf E
    &= -\dot{\widetilde{\mathbf B}},
    \qquad &\nabla\cdot \widetilde{\mathbf B}
    = 0, \label{eq:MaxwellCommon1}\\
    \nabla\widetilde{^\times} \widetilde{\mathbf H}
    &= \dot{\mathbf D} + \mathbf j,
    \qquad &\nabla\cdot\mathbf D
    = \rho, \label{eq:MaxwellCommon2}
\end{eqnarray}
accompanied, in the simplest case, by the constitutive relations
$\mathbf D = \epsilon \mathbf E$ and $\widetilde{\mathbf B} = \mu
\widetilde{\mathbf H}$\footnote{To end up with SI units, one puts
$\epsilon = \, \aob{\epsilon}{\mathrm r}\, \aob{\epsilon}{0}$ and
$\mu  = \, \aob{\mu}{\mathrm r} \, \aob{\mu}{0}$ with
$\aob{\epsilon}{0} \,= 10^7/(4\pi c^2)$ and $\aob{\mu}{0} \,= 4\pi
10^{-7}$.}. Dots over characters denote time derivatives; tildes
over $\mathbf H$ and $\mathbf B$ show explicitly that these
quantities are regarded as pseudo vectors (such notation follows
J.\,A.\,Schouten~\cite{SchoutenBOOK1951}); curls
$(\nabla\widetilde{^\times})^{ik} = \widetilde{\mathfrak E}^{ijk}
\partial_j$ are also defined through the pseudo density
$\widetilde{\mathfrak{E}}^{ijk}$, $i,j,k = 1,2,3$, equal to the
Levi-Civita permutation symbol $\varepsilon^{ijk}$ in any
coordinate system~\cite[p.~158]{LeviCivitaBOOK1927}:
\begin{equation}\label{eq:LeviCivitaAbs}
    \widetilde{\mathfrak E}^{i'j'k'} = \Delta^{-1}
    J_{i}^{i'} J_{j}^{j'} J_{k}^{k'} \widetilde{\mathfrak
    E}^{ijk}, \qquad \widetilde{\mathfrak
    E}^{ijk} \doteq \varepsilon^{ijk}.
\end{equation}
The $\doteq$ sign means equality in a given (here, in an arbitrary
right- or left-hand) coordinate system. With use of classic
`kernel-index' notation, we write Maxwell's equations
(\ref{eq:MaxwellCommon1}), (\ref{eq:MaxwellCommon2}) in a more
general form, valid assumably in arbitrary curvilinear,
nonorthogonal coordinates:
\begin{eqnarray}
    \widetilde{\mathfrak E}^{ijk}\partial_j E_k &=
    -\dot{\widetilde{\mathfrak B}}{}^i, \qquad
    &\partial_{i}\widetilde{\mathfrak B}^i = 0,
    \label{eq:MaxwellCommonInv1}\\
    \widetilde{\mathfrak E}^{ijk} \partial_j \widetilde{H}_k
    &= \dot{\mathfrak D}^i + \mathfrak{j}^i,
    \qquad &\partial_i \mathfrak{D}^i = \rho.
    \label{eq:MaxwellCommonInv2}
\end{eqnarray}
Here $E_i$ and $\widetilde{H}_i$ are the covariant vector and
pseudo vector; $\mathfrak{D}^i$ and $\widetilde{\mathfrak B}^i$
the contravariant vector density and pseudo density of weight
$+1$, as reflected by Gothic kernels; $\mathfrak{j}^i$ and $\rho$
the contravariant vector and scalar densities. This form follows
from the (here unquestioned) three-dimensional, generally
covariant Maxwell's equations~\cite{SchoutenBOOK1951,
PostBOOK1962}
\begin{eqnarray}
    2\partial_{[i} E_{j]} &= -\dot B_{ij}, \qquad
    \partial_{[i} B_{jk]} &= 0, \label{eq:MaxwellInv1}\\
    \partial_{j} \mathfrak{H}^{ij} &=
    \dot \mathfrak{D}^i + \mathfrak{j}^i, \qquad
    \partial_{i}\mathfrak{D}^{i} &= \rho, \label{eq:MaxwellInv2}
\end{eqnarray}
if one makes use of the dual equivalents
$\widetilde{\mathfrak{B}}^i = \frac12 \widetilde{\mathfrak
E}^{ijk} B_{jk}$ and $\widetilde{H}_i = \frac12
\widetilde{\mathfrak e}_{ijk} \mathfrak{H}^{jk}$, as suggested
e.g.\ by equations (2.35) and (2.37) in \cite{PostBOOK1962}. In
(\ref{eq:MaxwellInv1}) and (\ref{eq:MaxwellInv2}), the square
brackets denote alternation,  $B_{ij} = -B_{ji}$ is the covariant
bivector, and $\mathfrak{H}^{ij} = -\mathfrak{H}^{ji}$ the
contravariant bivector density. Assuming the constitutive
relations are $\widetilde{\mathfrak{B}}^i = \mu^{ij}
\widetilde{H}_j$ and $\mathfrak{D}^i = \epsilon^{ij} E_j$, one
concludes that the permittivity and permeability are tensor
densities of weight $+1$, so that they are transformed as
\begin{equation}\label{eq:EpsMuTransform}
    \epsilon^{i'j'} = |\Delta|^{-1}
    J_{i}^{i'} J_{j}^{j'} \epsilon^{ij},
    \qquad
    \mu^{i'j'} = |\Delta|^{-1}
    J_{i}^{i'} J_{j}^{j'} \mu^{ij}.
\end{equation}
An evident alternative to (\ref{eq:MaxwellCommonInv1}),
(\ref{eq:MaxwellCommonInv2}) can be constructed by using in
(\ref{eq:MaxwellInv1}) and (\ref{eq:MaxwellInv2}) the duals
$\mathfrak{B}^i = \frac{1}{2} \mathfrak{E}^{ijk} B_{jk}$ and $H_i
= \frac{1}{2} \mathfrak{e}_{ijk} \mathfrak{H}^{jk}$ instead, with
the ordinary densities $\mathfrak{E}^{ijk}$ and
$\mathfrak{e}_{ijk}$ equal to the Levi-Civita symbol for proper
transformations, and changing the sign for improper ones:
\begin{equation}\label{eq:LeviCivitaOrdinary}
    \mathfrak E^{i'j'k'} = |\Delta|^{-1}
    J_{i}^{i'} J_{j}^{j'} J_{k}^{k'} \mathfrak E^{ijk},
    \qquad \mathfrak E^{ijk} \doteq \varepsilon^{ijk}.
\end{equation}
Here $\doteq$ stands for equality in an arbitrary but right-hand
coordinate system. We see no reason to agree that the duals
constructed with $\mathfrak E^{ijk}$ and $\mathfrak{e}_{ijk}$
``are only meaningful for proper
transformations''~\cite[p.~41]{PostBOOK1962}. An appropriate form
of the generally covariant Maxwell's equations~--- compare to
(\ref{eq:MaxwellCommonInv1}), (\ref{eq:MaxwellCommonInv2})~---
reads
\begin{eqnarray}
    \mathfrak{E}^{ijk}\partial_j E_k &=
    -\dot \mathfrak{B}^i, \qquad
    &\partial_{i}\mathfrak{B}^i = 0,
    \label{eq:MaxwellMyInv1}\\
    \mathfrak{E}^{ijk}\partial_j H_k
    &= \dot \mathfrak{D}^i + \mathfrak{j}^i,
    \qquad &\partial_{i} \mathfrak D^i = \rho.
    \label{eq:MaxwellMyInv2}
\end{eqnarray}
Given the constitutive relations are $\mathfrak{B}^i = \mu^{ij}
H_j$ and $\mathfrak{D}^i = \epsilon^{ij} E_j$, the transformation
laws for the permittivity and permeability are exactly the same as
given by (\ref{eq:EpsMuTransform}). In standard three-vector form,
(\ref{eq:MaxwellMyInv1}), (\ref{eq:MaxwellMyInv2}) are reduced to
the commonly known equations (\ref{eq:MaxwellCommon1}),
(\ref{eq:MaxwellCommon2}) except for the tildes:
\begin{eqnarray}
    \nabla \times \mathbf E
    &= -\dot{\mathbf B},
    \qquad &\nabla\cdot \mathbf B
    = 0, \label{eq:MaxwellMyCommon1}\\
    \nabla \times \mathbf H
    &= \dot{\mathbf D} + \mathbf j,
    \qquad &\nabla\cdot\mathbf D
    = \rho. \label{eq:MaxwellMyCommon2}
\end{eqnarray}
Two differences from the standard formulation thus is that all
vectors are `polar', and the curl $\nabla\times\mathbf F =
\mathfrak{E}^{ijk} \partial_j F_k$. This seems perfectly sound
were we to assign any `rigidity' to the picture of magnetic `field
lines' for stationary media, instead of absolutizing the cross
product definition by use of pseudo permutation objects.

\section{Equivalence in real space}\label{sec:illustrations}

Most often in formulas that define observable (directly
measurable) quantities in terms of $\mathbf H$ or $\mathbf B$,
these latter fields enter in some conjunction with the vectorial
product operation~--- like $\nabla\times\mathbf H$ in Ampere's
law, $\mathbf j\times\mathbf B$ in the expression for the Lorentz
force density, or $\mathbf H^\times$ in the terms in constitutive
relations accounting for induced optical activity. If we note that
the neighboring tildes, if present, one over cross product and
other over the $\mathbf H$ or $\mathbf B$ vector, do always
annihilate, we conclude that the `polar-axial' and `pure polar'
formulations of electrodynamics are largely equivalent.

Now we also show the equivalence of the two formulations in
classifying the eigenmodes of symmetric systems by symmetry
arguments. In the following, consider reflection with respect to
the $x^1 = 0$ surface, $x^{1'} = -x^1$. The corresponding Jacobian
is $J^{i'}_{i} = \mathrm{diag}(-1, 1, 1) = J^{i}_{i'}$, its
determinant $\Delta = -1$. It is well known that reflection
symmetry of the system, $\epsilon'(\mathbf{x}') =
\epsilon(\mathbf{x})$ and $\mu'(\mathbf{x}') = \mu(\mathbf{x})$,
allows to replace the whole system by its half, with the PEC (the
`electric wall') or PMC (the `magnetic wall') boundary standing in
place of the reflection symmetry plane.

We write (\ref{eq:MaxwellMyInv1}), (\ref{eq:MaxwellMyInv2}) in
terms of covariant electric and magnetic vectors:
\begin{eqnarray}
    \mathfrak{E}^{ijk}\partial_j E_k &=
    -\mu^{ij} \dot H_j , \qquad
    &\partial_{i} \mu^{ij}H_j = 0,
    \label{eq:MaxwellMyEHInv1}\\
    \mathfrak{E}^{ijk}\partial_j H_k
    &= \epsilon^{ij} \dot E_j + \mathfrak{j}^i,
    \qquad &\partial_{i} \epsilon^{ij} E_j = \rho.
    \label{eq:MaxwellMyEHInv2}
\end{eqnarray}
In the `pure polar' formalism, both vectors are transformed under
reflection as
\begin{eqnarray}
E_{i'}(\mathbf{x}') &= J^{i}_{i'}E_{i}(J^{i'}_{i}x^i)\nonumber \\
&= \{-E_1(-x^1,x^2,x^3), E_2(-x^1,x^2,x^3),
E_3(-x^1,x^2,x^3)\},\label{eqEreflect}\\
H_{i'}(\mathbf{x}') &= J^{i}_{i'}H_{i}(J^{i'}_{i}x^i)\nonumber \\
&= \{-H_1(-x^1,x^2,x^3), H_2(-x^1,x^2,x^3),
H_3(-x^1,x^2,x^3)\}.\label{eqHreflect}
\end{eqnarray}
At the same time, if material objects are invariant under the
given transform, so that $\epsilon^{i'j'}(\mathbf{x}') =
\epsilon^{ij}(\mathbf{x})$ and $\mu^{i'j'}(\mathbf{x}') =
\mu^{ij}(\mathbf{x})$, then Maxwell's equations
(\ref{eq:MaxwellMyEHInv1}) and (\ref{eq:MaxwellMyEHInv2}) remain
unchanged except for the $-1$ multipliers acquired by the
left-hand sides, owing to the transformation rule for
$\mathfrak{E}^{ijk}$ (\ref{eq:LeviCivitaOrdinary}). Thus the
admissible new solutions (in the absence of free charges) are
$E_{i'}(\mathbf{x}') = \pm \alpha E_i(\mathbf{x})$ and
$H_{i'}(\mathbf{x}') = \mp \alpha H_i(\mathbf{x})$. Invariance of
the physical system (material objects plus electromagnetic fields)
to double reflection gives $\alpha^2 = 1$. Equating
(\ref{eqEreflect}) with $E_i$ and (\ref{eqHreflect}) with $-H_i$
yields
\numparts
\begin{eqnarray}
\fl \quad -&E_1(-x^1,x^2,x^3) = E_1(x^1,x^2,x^3),
\quad -&H_1(-x^1,x^2,x^3) = -H_1(x^1,x^2,x^3), \\
\fl &E_2(-x^1,x^2,x^3) = E_2(x^1,x^2,x^3),
\quad &H_2(-x^1,x^2,x^3) = -H_2(x^1,x^2,x^3), \\
\fl &E_3(-x^1,x^2,x^3) = E_3(x^1,x^2,x^3), \quad
&H_3(-x^1,x^2,x^3)= -H_3(x^1,x^2,x^3).
\end{eqnarray}
\endnumparts
These conditions define even modes with the `numerical' PMC plane
at $x^1 = 0$: $E_1|_{x^1 = 0} = 0$, $H_2|_{x^1 = 0} = 0$,
$H_3|_{x^1 = 0} = 0$ (the `physical' perfect conductor planes are
not so readily expressed in nonorthogonal coordinates). The
conditions obtained in a similar way with $\alpha = -1$ define the
numerical PEC boundary and odd modes. Thus, the `pure polar'
Maxwell's equations (\ref{eq:MaxwellMyInv1}),
(\ref{eq:MaxwellMyInv2}) lead to exactly the same mode
classification in optical waveguides and resonators exhibiting
reflection symmetry as does the standard `polar-axial'
formulation.

\section{Complex-coordinate scaling}\label{sec:PMLs}

In computational electrodynamics, the dominating way for treating
unbounded problems in the finite-difference time-domain (FDTD),
finite-difference frequency-domain, or finite-element calculations
is with the perfectly matched layer (PML) method. Introduced
originally in the `split-field' formulation by
J.\,P.\,B\'erenger~\cite{Berenger94} and soon afterwards recast in
the `uniaxial'~\cite{Sacks95,Gedney96} and
stretched-coordinate~\cite{Chew94, Rappaport95, Petropoulos98,
Teixeira98, Lalanne05} forms, the PML concept stands as ``one of
the most significant advances in the historical development of the
FDTD method''~\cite{TafloveBOOK98}.

The method is closely related to complex-coordinate transforms in
classical quantum theory of atomic
resonances~\cite{Moiseyev98}~--- though this is rarely appreciated
by the electromagnetics modelling community. In electromagnetics,
the method is implemented with surprising ease. Without loss of
generality, we consider domain truncation in one, $x^1\equiv x$,
direction. For media described by diagonal $\epsilon$ and $\mu$
matrices in the given coordinates, those matrices are modified in
the PML regions according to
\begin{equation}\label{eq:PML1d}
    \hat\epsilon = \epsilon\aob{\Lambda}{x}, \qquad
    \hat\mu = \mu\aob{\Lambda}{x}, \qquad
    \aob{\Lambda}{x} \, = \left(\begin{array}{ccc}
  1/s_x & 0 & 0 \\
  0 & s_x & 0 \\
  0 & 0 & s_x \\
    \end{array}\right)
\end{equation}
with the complex function
\begin{equation}\label{eq:sx}
    s_x(x',\omega) = \zeta(x') + \mathrm i\frac{\sigma(x')}{\omega}
\end{equation}
where $x'\in[-b,b]$ is the bounded computation-space coordinate,
$\zeta(x')$ and $\sigma(x')$ are real-valued functions equal to 1
and, respectively, 0 over $[-a,a]\subset[-b,b]$, while
$\sigma(x')>0$ for $a<|x'|\leq b$ in order to damp the
$\exp[\mathrm i(kx'-\omega t)]$ oscillatory waves inside the PML
regions. Odd-power frequency dependence is included in the
imaginary part of (\ref{eq:sx}), and hence in (\ref{eq:PML1d}),
for causality reasons~\cite[\S123]{LandauLifshitzCOURSEstat}. It
is repeatedly claimed that the structure of permittivity and
permeability modified within the PMLs according to
(\ref{eq:PML1d}) can be recovered by complex coordinate scaling
\begin{equation}\label{eq:complexScaling}
    x = \int_0^{x'} s_x(x',\omega)\,\mathrm{d}x'.
\end{equation}
It is easy to see however, that only the `pure polar' formulation
(\ref{eq:MaxwellMyInv1}), (\ref{eq:MaxwellMyInv2}) leads to
(\ref{eq:PML1d}). In contrast, (\ref{eq:MaxwellCommonInv1}) and
(\ref{eq:MaxwellCommonInv2}) yield
\begin{equation}\label{eq:PML1dwrong}
    \aob{\Lambda}{x} \, = \left(\begin{array}{ccc}
  |s_x|/s_x^2 & 0 & 0 \\
  0 & |s_x| & 0 \\
  0 & 0 & |s_x| \\
    \end{array}\right)
\end{equation}
which is reduced to (\ref{eq:PML1d}) for real-valued coordinate
squeezing~\cite{ShyrokiIEEELett06}, but essentially differs from
(\ref{eq:PML1d}) given the $s_x$ function complex.

\begin{figure}
  \includegraphics[width=0.5\columnwidth]{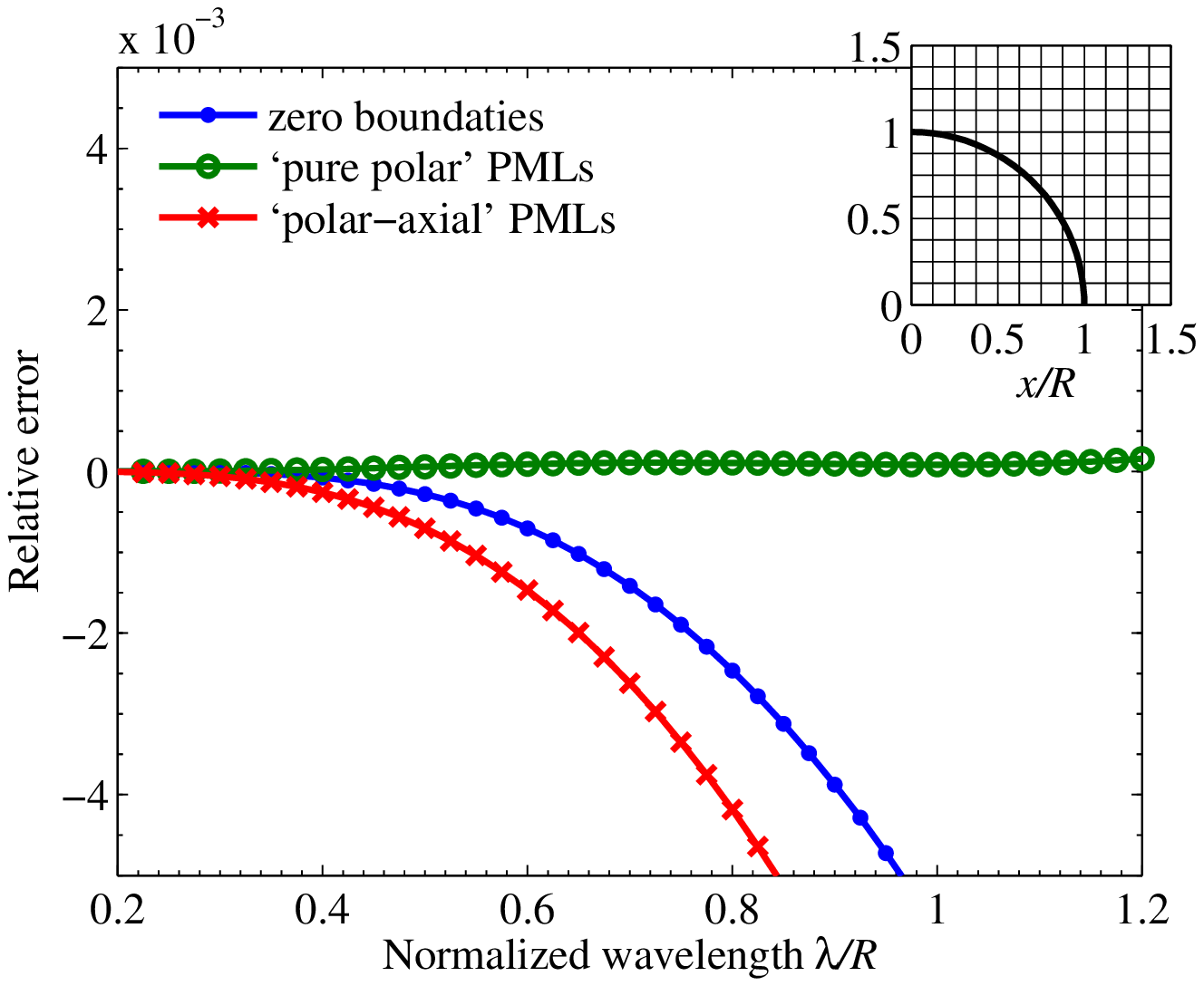}%
  \includegraphics[width=0.5\columnwidth]{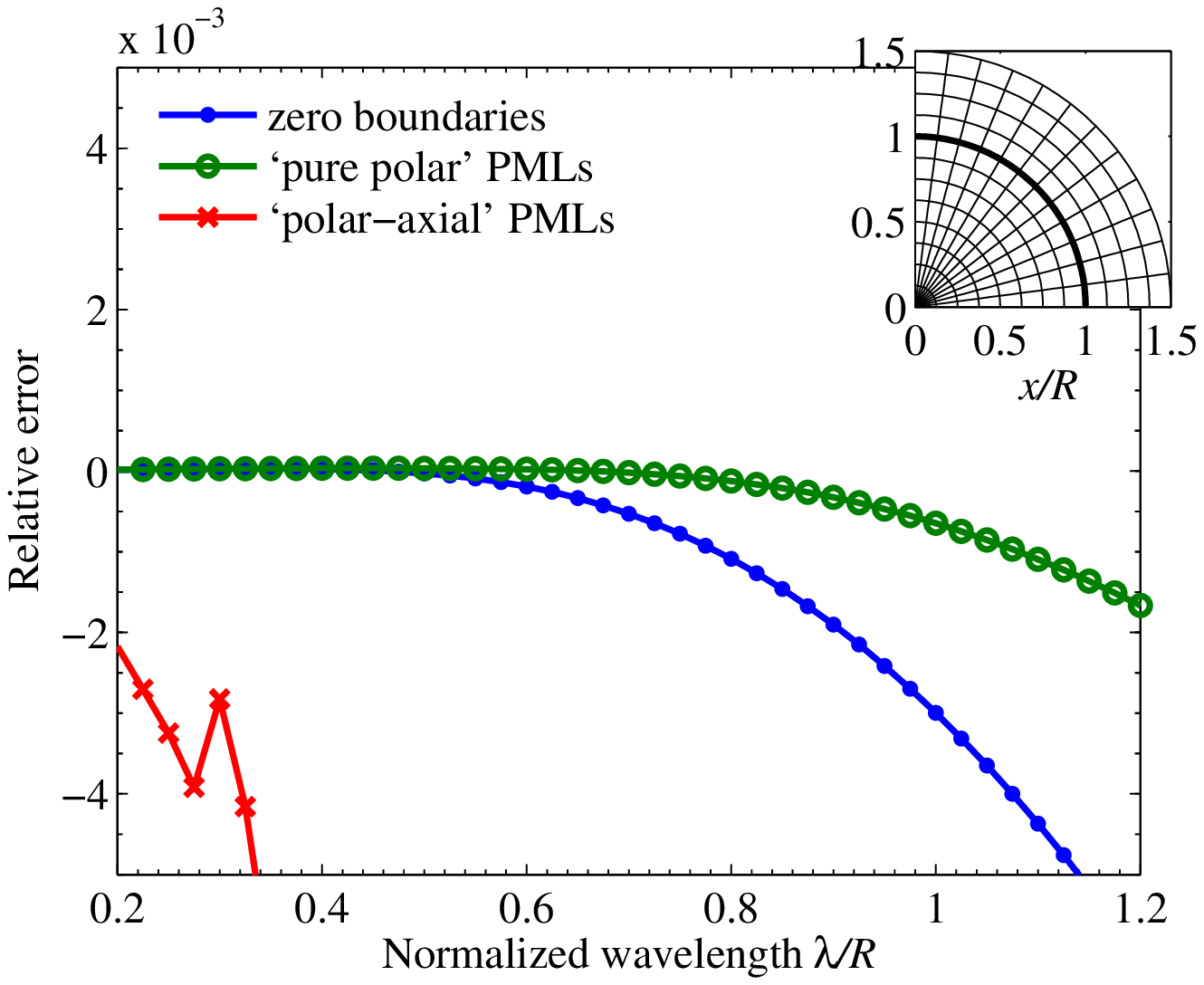}\\
  \caption{Mode index errors on Cartesian (left) and polar (right)
  grids. The superiority of the standard perfectly matched
  layers compatible with the `pure polar' Maxwell's equations
  (\ref{eq:MaxwellMyInv1}), (\ref{eq:MaxwellMyInv2}) is
  evident.}\label{fig:PMLpolaraxial}
\end{figure}
In order to compare the performance of (\ref{eq:PML1dwrong}) and
(\ref{eq:PML1d}), we simulated the fundamental mode of a
step-index fiber ($n_\mathrm{core}/n_\mathrm{clad} = 1.515/1.5$ as
in~\cite{ShyrokiIEEELett06, Tsuji00}) with the two-dimensional
finite-difference frequency-domain method on Cartesian and polar
grids. In figure~\ref{fig:PMLpolaraxial}, the fundamental mode
index errors are plotted, associated with the standard (`pure
polar' consistent) PMLs, the `polar-axial' PMLs, and simple zero
boundaries. The corresponding physical domains are $[0\dots 1.5R]
\times [0\dots 1.5R]$ and $[0\dots 1.5R] \times [0\dots\pi/2]$,
with resolution $40\times 40$ and $40\times 20$ pixels,
respectively. 

\section{Summary}\label{sec:conclusions}

Standard (`polar-axial') and alternative (`pure polar')
formulations of Maxwell's equations were presented in
\sref{sec:Maxwell}. In most cases, as illustrated in
\sref{sec:illustrations}, both approaches yield equivalent
description of electromagnetics phenomena. In \sref{sec:PMLs} we
pointed however that it is the `pure polar' formalism that
complies with the PML method used with much success in
computational electrodynamics. An interesting question is, To
which extent can the classic electric and magnetic fields~---
whose parity is a matter of controversy~--- be associated with the
`electric $2^j$-pole' and `magnetic $2^j$-pole' photon states of
different parity in quantum electrodynamics?


\end{document}